\documentclass[useAMS,usenatbib,a4paper]{mn2e}

\usepackage{amssymb}
\usepackage{psfrag}
\usepackage{graphicx}
\usepackage{subfigure}
\usepackage{booktabs}
\usepackage{txfonts}
\usepackage{ifpdf}
\usepackage{hyperref}

\newif\ifbw
\bwfalse

\newcommand{\cmb}{{CMB}}
\newcommand{\cmbtext}{{cosmic microwave background}}
\newcommand{\bianchi}{{Bianchi}}
\newcommand{\bianchiviih}{{Bianchi VII{$_{\lowercase{\rm h}}$}}  }
\newcommand{\healpix}{{\tt HEALPix}}

\newcommand{\lambdaarch}{{LAMBDA}}
\newcommand{\lambdaarchtext}{{Legacy Archive for Microwave Background Data Analysis}}
\newcommand{\wmap}{{WMAP}  }
\newcommand{\wmaptext}{{Wilkinson Microwave Anisotropy Probe}}
\newcommand{\ilc}{{ILC}}
\newcommand{\ilctext}{{internal linear combination}}

\newcommand{\eqn}[1]{(#1)}

\newcommand{\fig}[1]{Fig.~#1}

\title[\bianchiviih\ models and the cold spot texture]
  {\bianchiviih\ models and the cold spot texture}

\author[M. Bridges et al.]
  {M.~Bridges$^{1}$\thanks{E-mail: m.bridges@mrao.cam.ac.uk}, J.~D.~McEwen$^{1}$, M.~Cruz$^{2}$, M.~P.~Hobson$^{1}$, A.~N.~Lasenby$^{1}$,
   \newauthor P.~Vielva$^{2}$, E.~Mart\'inez-Gonz\'alez$^{2}$\\ 
  $^1$Astrophysics Group, 
      Cavendish Laboratory,  J.~J.~Thomson Avenue,
      Cambridge CB3 0HE, UK\\
  $^2$Instituto de F\'{\i}sica de Cantabria,
      {CSIC-Universidad de Cantabria}, Avda.\ los Castros s/n,
      39005, Santander, Spain\\
}

\date{Accepted ---. Received ---; in original form \today}
\pagerange{\pageref{firstpage}--\pageref{lastpage}} 
\pubyear{2007}

\def\LaTeX{L\kern-.36em\raise.3ex\hbox{a}\kern-.15em
    T\kern-.1667em\lower.7ex\hbox{E}\kern-.125emX}

\begin{document}
\label{firstpage}
\maketitle


\begin{abstract}
We have returned to our previous \bianchiviih analysis in light of the \citet{cruz:2007}
suggestion that the cold spot observed near the southern Galactic pole may be 
a remnant temperature perturbation of a cosmic texture. In \citet{bridges:2006b}
we found two favoured left handed \bianchiviih templates with restricted prior probabilities
so that the template was centred close to the cold spot. 
Using \wmap data `corrected' for the texture fit we have now reexamined 
both models to assess any changes to these conclusions. We find that both models 
are left almost entirely unconstrained by the data and consequently exhibit 
significantly reduced Bayesian evidences. Both models are now \emph{disfavoured} 
by the data.  
This result reinforces our previous suggestion that the cold spot 
was likely to be driving any \bianchiviih detection. 
\end{abstract}



\section{Introduction}
\label{sec:introduction}

The anomalous cold spot discovered \citep{vielva:2004,cruz:2005,cruz:2006a,cruz:2006b} in \wmaptext\ (\wmap)
observations of the \cmbtext\ (\cmb) \citep{bennett:2003a,hinshaw:2006} has been shown by \citet{cruz:2007} to be
consistent with a temperature perturbation induced by a cosmic texture. 
A cosmic texture is a particular type of cosmic defect predicted by certain
theories of high energy physics. They form at symmetry breaking phase
transitions in the early Universe and are extremely energetic events
producing both hot and cold spots in the CMB.
 The texture hypothesis is the most
convincing explanation of the cold spot anomaly made to date, 
however evidence for the texture requires further observational support. 
One of the most promising aspects of the texture hypothesis is that tests may be performed using
future observations of \cmb\ polarisation, which may either substantiate or refute evidence for the presence of a
texture.  The discovery of a texture would
have such profound implications for our understanding of the early Universe that the cold spot-texture hypothesis
certainly warrants further investigation.

\bianchiviih\ models in which the universe exhibits a global rotation and shear have also
been considered in an attempt to explain many of the anomalies in the \wmap\ data.  A
positive detection of a \bianchiviih\ template embedded in the \wmap\ data was first made by
\citet{jaffe:2005}.  After `correcting' the \wmap\ data for this template a number of
previously reported anomalies in the data disappear
\citep{jaffe:2005,lm:2006,cayon:2006,mcewen:2006:bianchi}.  However, the corresponding
best-fit \bianchiviih\ template was shown to be incompatible with concordance cosmology
\citep{jaffe:2006b}.  \citet{bridges:2006b} performed a more rigorous Bayesian MCMC analysis
to determine the evidence for \bianchiviih\ models, concluding that there is weak evidence
\emph{against} a \bianchiviih\ component when the axis of the \bianchi\ coordinate system
is allowed to vary over the entire sky.  However, weak evidence \emph{for} a Bianchi
template remains when the axis is restricted to lie in the direction of the cold spot. 
These results suggest that it may have been the cold spot that was driving the positive
template detection made by \citet{jaffe:2005}.

The focus of this work is to determine whether there is evidence for any
\bianchiviih\ component embedded in the \wmap\ data once the data is
`corrected' for the cold spot texture template determined by
\citet{cruz:2007}.  If all positive evidence vanishes, and one accepts the
texture explanation of the cold spot, then \bianchiviih\ models may be
rejected definitively.  The remainder of this letter is structured as
follows.  Firstly, the procedure followed by \citet{cruz:2007} to fit a
texture template to the cold spot in the combined, foreground-cleaned Q-V-W
map (hereafter \wmap\ co-added map) is discussed, before the template is used
to `correct' the \wmap\ \ilctext\ (\ilc) map (since the subsequent MCMC
analysis requires a full-sky map and the co-added map necessitates a Galactic
cut). Using different processed versions of the \wmap\ data for template
fitting and analysis is acceptable and was shown by
\citet{mcewen:2006:bianchi} not to alter analysis results for the \bianchi\
case.  Secondly, the Bayesian evidence for the texture `corrected' \wmap\
\ilc\ map is computed and compared to the evidence computed previously
\citep{bridges:2006b} for the original data.  Concluding remarks are then
made.

\section{Cold spot texture fitting}
\label{sec:spot}

The cold spot has been shown by \citet{cruz:2007} to be consistent with a temperature perturbation induced in the CMB by cosmic texture at redshift $z \approx 6$.  Unwinding events associated with texture induce cold and hot spots in the CMB.  An analytic approximation for the temperature profile produced by a texture is given by \citep{turok:1990}
\begin{equation}
\frac{\delta T}{T} = \frac{\pm \epsilon}{\sqrt{1 + 4 ( \theta / \theta_{\rm C} ) ^2}},
\label{eqn:texture}
\end{equation}
where $\theta$ denotes angular separation, $\theta_{\rm C}$ is the scale parameter of the texture and $\epsilon$ is the amplitude parameter 
related to the symmetry breaking scale $\phi_0$.
This temperature profile is an approximation and is not valid for large co-moving scales.  Consequently, for practical applications the profile is truncated beyond its half-maximum by matching its value and derivative to a Gaussian, as discussed by \citet{magueijo:1995}.  $\theta_{\rm C}$ is then equal to the standard deviation of the matching Gaussian.  The resulting texture temperature profile was fitted to the cold spot by \citet{cruz:2007} and found to be favoured to the null hypothesis of no texture.  In this section we review the texture fitting procedure performed by \citet{cruz:2007} and present the best-fit texture profile determined.

The co-added map of the three-year \wmap\ data release \citep{hinshaw:2006} is used to fit the texture profile to the cold spot.  \citet{cruz:2007} degrade the co-added map in the HEALPix\footnote{\url{http://healpix.jpl.nasa.gov/}} pixelization scheme \citep{gorski:2005} 
to a resolution parameter of $N_{\rm side} = 64$.  This resolution is sufficient to retain all information associated with the
cold spot, which occurs on a half-width scale of $\sim5^\circ$, and reduces the number of pixels used in the template fitting,
thereby reducing the computational cost of the fitting procedure.
The template fitting was performed in a circular area of $20^\circ$ radius centred at Galactic coordinates
($b = -57^\circ, l = 209^\circ$). 
Although the total angular size of the cold spot is $\sim10^\circ$, it is necessary to consider a $20^\circ$ radius
patch in order to take into account the entire neighbourhood of the spot since the original Spherical Mexican Hat Wavelet analyses
that highlighted the spot \citep{vielva:2004} convolves all pixels in this region.  These pixels could therefore contribute in an important way to the detected structure and must be included when fitting the texture template.

A hybrid Bayesian-frequentist approach is considered for the template fitting performed by \citet{cruz:2007}.  The Bayesian evidence ratio is used to perform hypothesis testing, which is then calibrated using Monte Carlo simulations.  The data are found to favour the alternative hypothesis $H_{\rm a}$ that a texture is present. 
Parameters $\epsilon$ and $\theta_{\rm C}$ of the temperature template profile $\mathbf{T}$, defined by \eqn{\ref{eqn:texture}}, are then determined by maximising the posterior probability for the alternative hypothesis, where the posterior is given by Bayes' Theorem:
\begin{equation}
Pr({\mathbf \Theta} | {\mathbf D}, H_{\rm a}) \propto Pr( {\mathbf D} | {\mathbf \Theta}, H_{\rm a}) Pr( {\mathbf \Theta}| H_{\rm a}),
\label{eqn:bayes}
\end{equation} 
with likelihood $Pr(\mathbf{D}|\mathbf{\Theta},H_{\rm a})$ and prior $Pr(\mathbf{\Theta}|H_{\rm a})$, where $\mathbf{D}$ represents the WMAP co-added data and $\mathbf{\Theta}=(\epsilon,\theta_{\rm C})$.
The likelihood function is assumed to be Gaussian:
\begin{equation}
L \propto e^{-\frac{\chi^{2}}{2}},
\end{equation}
where
\begin{equation}
\chi^2 = (\mathbf{D}-\mathbf{T})^{\rm T} \mathbf{N}^{-1} (\mathbf{D} - \mathbf{T}),
\end{equation}
with the generalised noise covariance matrix $\mathbf{N}$ including both \cmb\ and noise contributions. 
The calculation of the noise contribution to $\mathbf{N}$ is straightforward since the noise in the \wmap\ co-added map is uncorrelated and well defined through the number of observations per pixel.
In order to obtain the \cmb\ contribution to the matrix, the covariance function for the \wmap\ co-added map is calculated taking into account both pixel and beam effects.
As a complementary test the \cmb\ covariance matrix is calculated through 70,000 Gaussian simulations.
\citet{cruz:2007} compared the $\chi^2$ values computed from the data to those obtained using simulations and found errors were negligible.
As a conservative prior on $\epsilon$, \citet{cruz:2007} choose \mbox{$0 \leq \epsilon \leq 10^{-4}$}, the COBE-normalised amplitude \citep{pen:1994,durrer:1999}.
A scale-invariant distribution of texture spots on the sky is predicted
\begin{equation}
\frac{dN_{spot}}{d \theta_C} = {8 \pi \nu \over 3}  \frac{\kappa^3}{\theta_C^3},
\label{eqn:dn_spots}
\end{equation}
where $\nu$ is a dimensionless constant and $\kappa$ a fraction of unity.
In order to obtain the prior for the scale parameter $\theta_{\rm C}$, \eqn{\ref{eqn:dn_spots}} is
normalised to unity between $\theta_{\rm min}$ and $\theta_{\rm max}$. Photon diffusion would
smear out textures smaller than a degree or so, hence the lower $\theta_{\rm C}$ bound is set to
$\theta_{\rm min} = 1^\circ$, according to the
resolution considered. At large scales textures are rare, hence the upper $\theta_{\rm C}$
bound is set to $\theta_{\rm max} = 15^\circ$. In this setting \citet{cruz:2007} obtain texture
parameter estimates of $\epsilon = 7.7\times 10^{-5}$ and $\theta_{\rm C} = 5.1^\circ$.  The
original co-added data, the best-fit texture template and their difference are shown on a small
patch on the sky in Fig 1. of \citet{cruz:2007}. 


\section{Evidence for Bianchi VII{$_{\lowercase{h}}$} models}
\label{sec:bianchi}

\bianchiviih models induce characteristic `spiral' temperature fluctuations on the CMB sky. They are
described by four independent quantities: a matter energy density $\Omega_{\rm m}$, 
a dark energy density $\Omega_{\Lambda}$, the current vorticity $\omega$ and $h$, which physically 
relates the characteristic wavelength over which the principle axes of shear and vorticity change orientation, 
and determines
the `tightness' of the spiral pattern. The pattern position is defined by Euler angles $\alpha$, $\beta$ and $\gamma$ 
\footnote{We adopt the active $zyz$ Euler convention corresponding to the rotation of a physical body
in a {\it{fixed}} coordinate system about the $z$, $y$ and $z$ axes by $\gamma$, $\beta$ and $\alpha$
respectively.}. Additionally one must specify the direction of rotation or handedness of the spiral.

\citet{jaffe:2005} found evidence of a significant correlation between a
\bianchiviih model and the \wmap 1-year data. In \citet{bridges:2006b} we performed a more rigorous
Bayesian analysis that confirmed the \citet{jaffe:2005} fit in \wmap 3-year data. In this
framework our entire inference is contained in the multidimensional posterior distribution
from which we can extract marginalised parameter constraints \emph{and} the comparative Bayesian 
evidence to select the most appropriate model parameterisation. We aimed to use the evidence to 
establish whether it was necessary to include a \bianchiviih component in addition to a standard $\Lambda$CDM cosmology.
We concluded
that the only significant Bayesian evidence favouring such an inclusion existed 
where the central position of the \bianchiviih spiral was fixed close to the cold spot. In this letter
we aim to establish how the significance of this conclusion has changed in light of the texture
fit of \citet{cruz:2007}. We use the \wmap 3-year ILC map (\fig{\ref{fig:maps}} (a)) subtracting the
cold spot template described by \citet{cruz:2007} (\fig{\ref{fig:maps}} (c)) to yield the `corrected' map
shown in \fig{\ref{fig:maps}} (b). 

The models we will consider here are those of A, D \& G from \citet{bridges:2006b}, for convenience we summarise the
parameter combinations and priors in Tables \ref{table:priors} \& \ref{table:models}. Model A is a standard
$\Lambda$CDM cosmology, with one free parameter $A_{\rm s}$, the amplitude of scalar fluctuations, this being the only
parameter constrained on these large scales.  Both models D and G had the Bianchi template
axis fixed close to the cold spot
via heavily constrained priors on $\alpha$ and $\beta$. However D and G differ in the inclusion of a dark energy
density component which significantly opens up the parameter space along a degeneracy in the $\Omega_{\Lambda} -
\Omega_{\rm m}$ plane (see Fig. 6 in \citealt{bridges:2006b}). It was first suggested by \citet{jaffe:2006b} that a
dark energy term might give the \bianchiviih models more freedom to accomodate a close to flat cosmology preferred by
\wmap and others. Although the Bianchi degeneracy found is intriguingly close  to the geometric degeneracy seen in
\citet{spergel:2007} it does not cross the \wmap confidence contours above the $2\sigma$ level. From this we concluded
that the set of cosmological  parameters preferred by \wmap are incompatible with those of any \bianchiviih component.
We proceeded in our analysis by  decoupling the \bianchiviih matter and energy densities from their $\Lambda$CDM
counterparts so that the \bianchiviih component was essentially treated as a template with energy densities
$\Omega_{\rm m}^{\rm B}$ and $\Omega_{\Lambda}^{\rm B}$.  Left-handed models D and G were the only ones found to show 
marginally significant evidence favouring their inclusion and so we have chosen these to carry over for this
analysis.  Morphologically both D and G are very similar (owing to the degeneracy described above); the best fitting
template is shown in  \fig{\ref{fig:maps}} (d). 

\begin{table}
\begin{center}
\caption{Summary of Bianchi ${\rm VII_h}$ component priors used in this analysis.}
\begin{tabular}{|c|}
    \hline
 \textbf{Full Bianchi}\\
    \hline
 $\Omega_{tot}^{\rm B} = [0.01,0.99]$\\  
 $\Omega_{m}^{\rm B} = [0.01,0.99]$\\
 $h = [ 0.01,1]$\\
 $\omega = [0 ,20 ]\times 10^{-10}$\\
 $\gamma = [ 0, 2\pi ]$${\rm rads}$\\
 Chirality = L/R\\
    \hline
\end{tabular}
\label{table:priors}
\end{center}
\end{table}

\begin{table}
\begin{center}
\caption{Cosmological and Bianchi parameterisations for each of the parameter subsets studied.}
\begin{tabular}{|c||c||c|}
    \hline
 \textbf{Model} &  \textbf{Cosmology} & \textbf{Bianchi}\\
    \hline
 A & $A_s$ & -\\
 D & $A_s$ & $\Omega_m^B$, $\Omega_{tot}^B$, $h$, $\omega$, $\gamma$, L/R\\
 G & $A_s$ & $\Omega_m^B$, $h$, $\omega$, $\gamma$, L/R\\
 \hline 
\end{tabular}
\label{table:models}
\end{center}
\end{table}

\newlength{\mapplotwidth}
\setlength{\mapplotwidth}{75mm}

\begin{figure}
\centering
\subfigure[WMAP ILC map]{\includegraphics[clip=,width=\mapplotwidth]{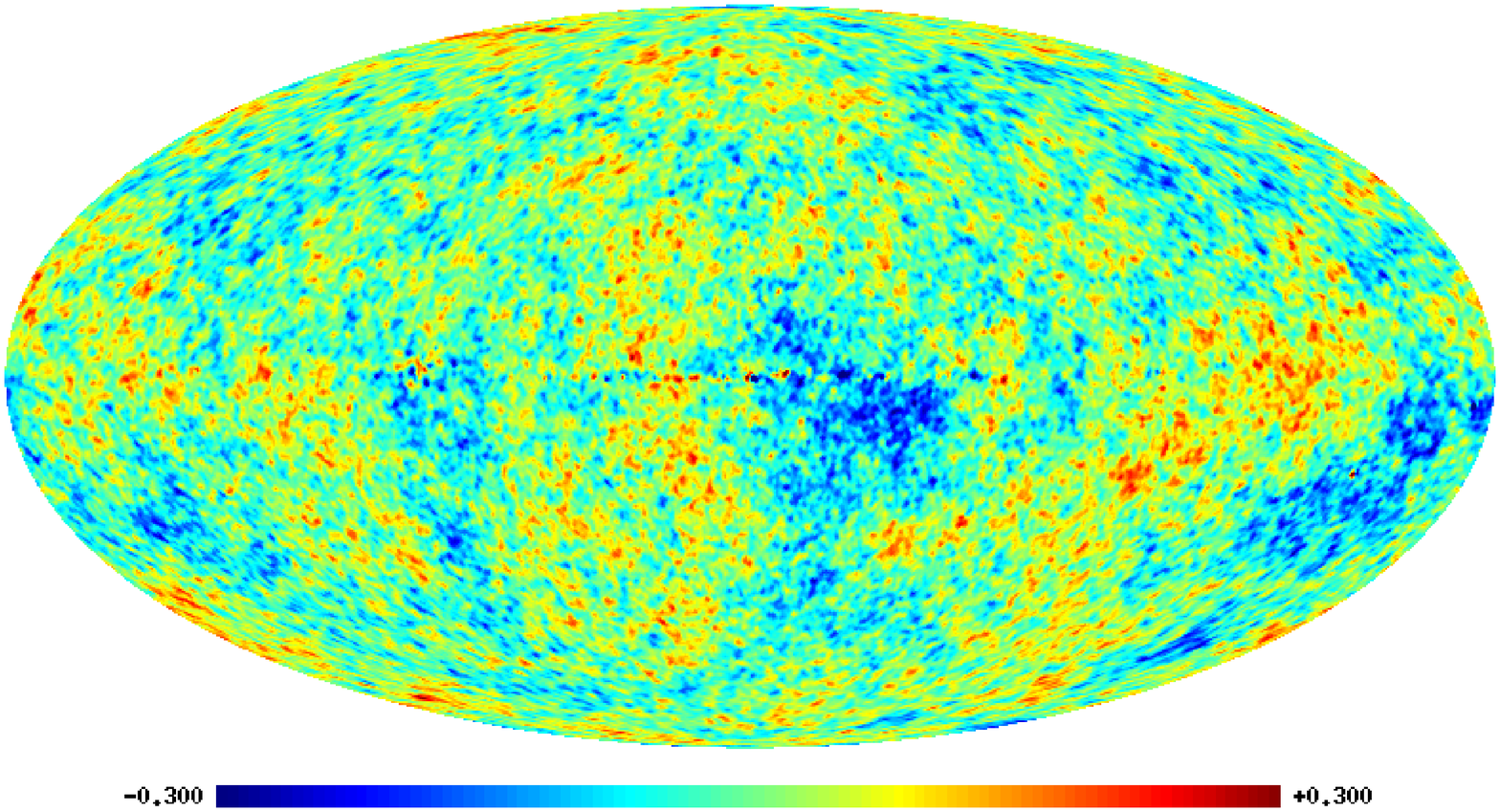}}
\subfigure[Cold spot corrected WMAP ILC map]{\includegraphics[clip=,width=\mapplotwidth]{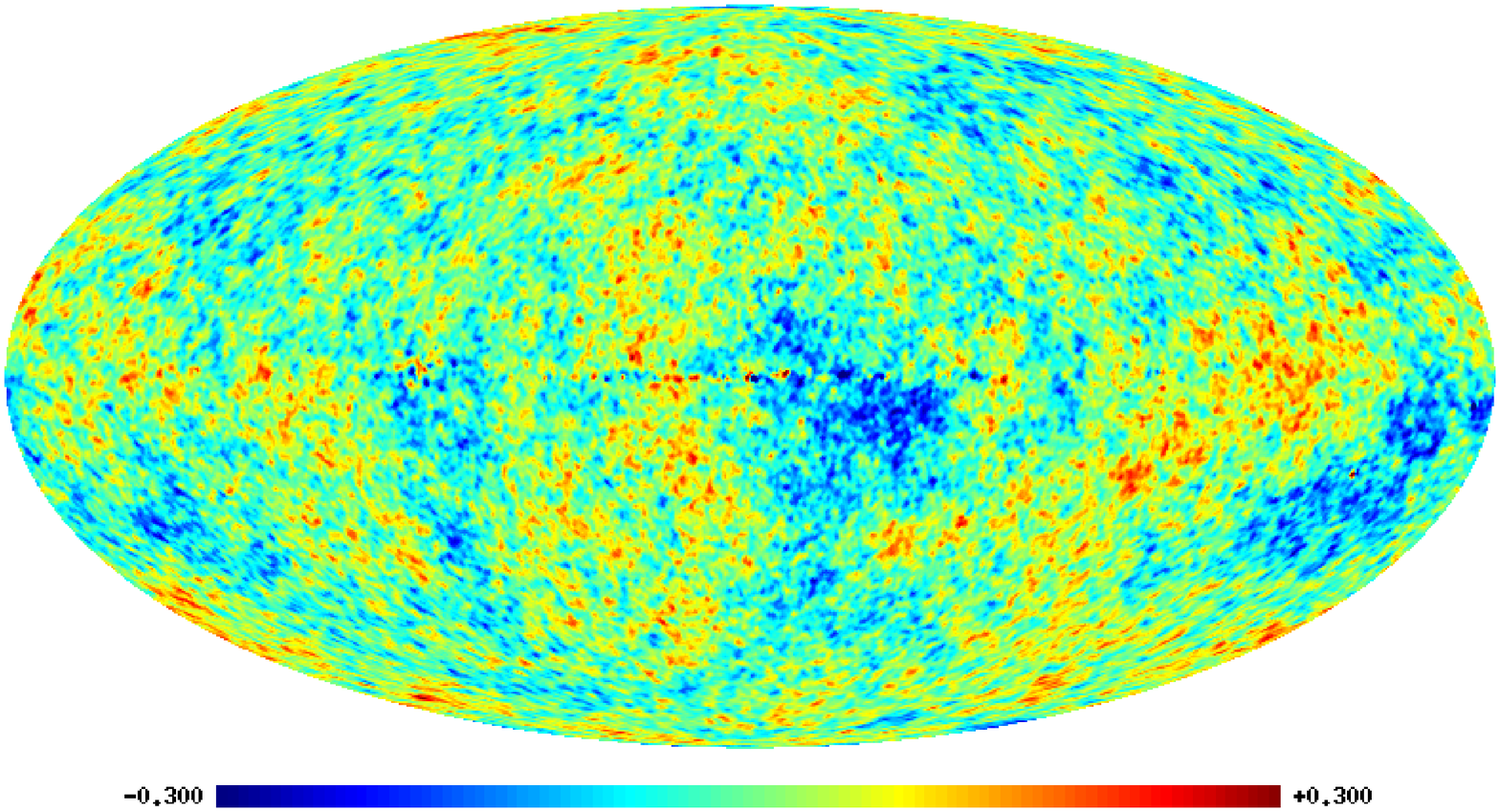}}
\subfigure[Cold spot template]{\includegraphics[clip=,width=\mapplotwidth]{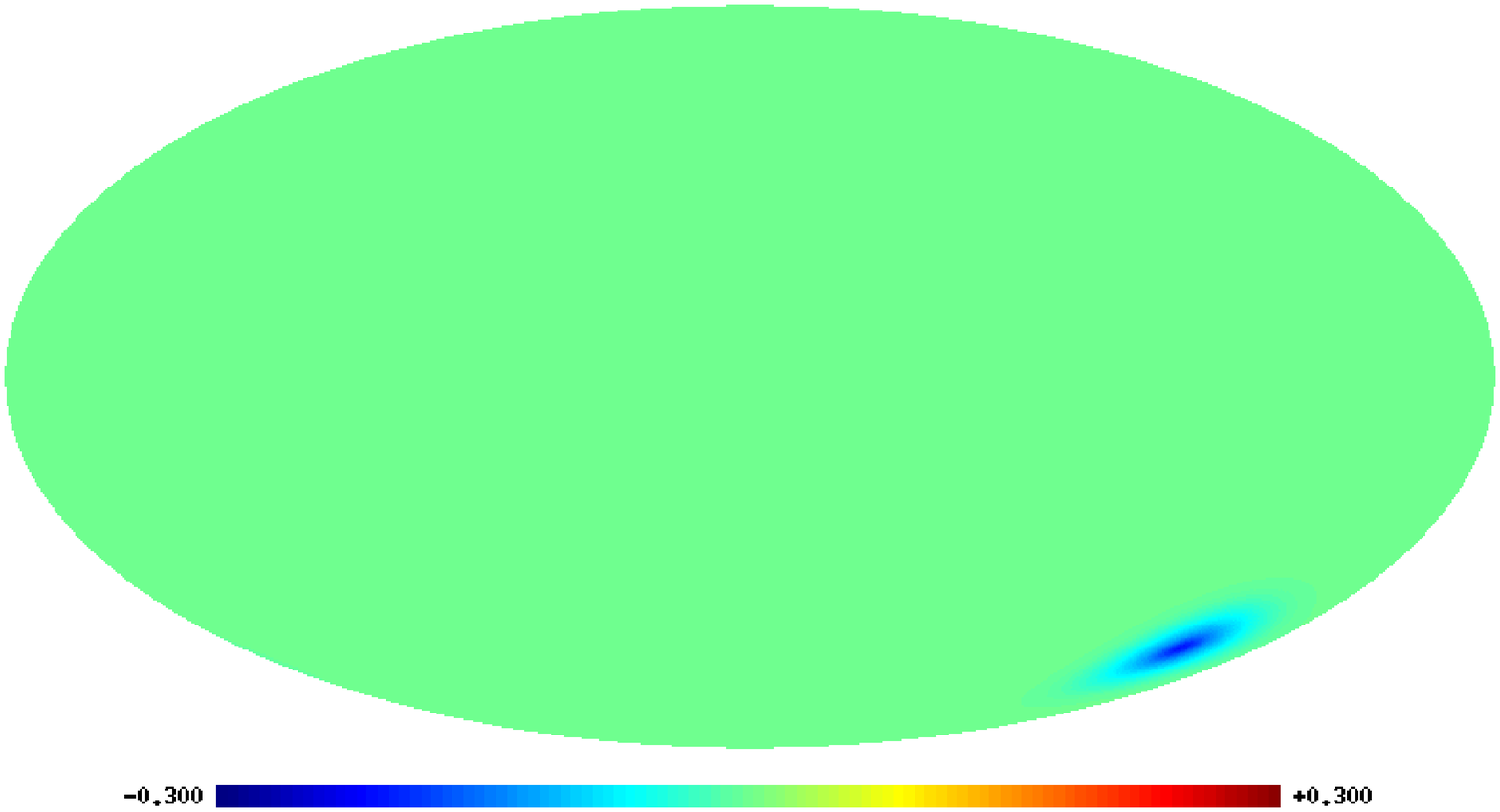}}
\subfigure[Previous best-fit Bianchi template]{\includegraphics[clip=,width=\mapplotwidth]{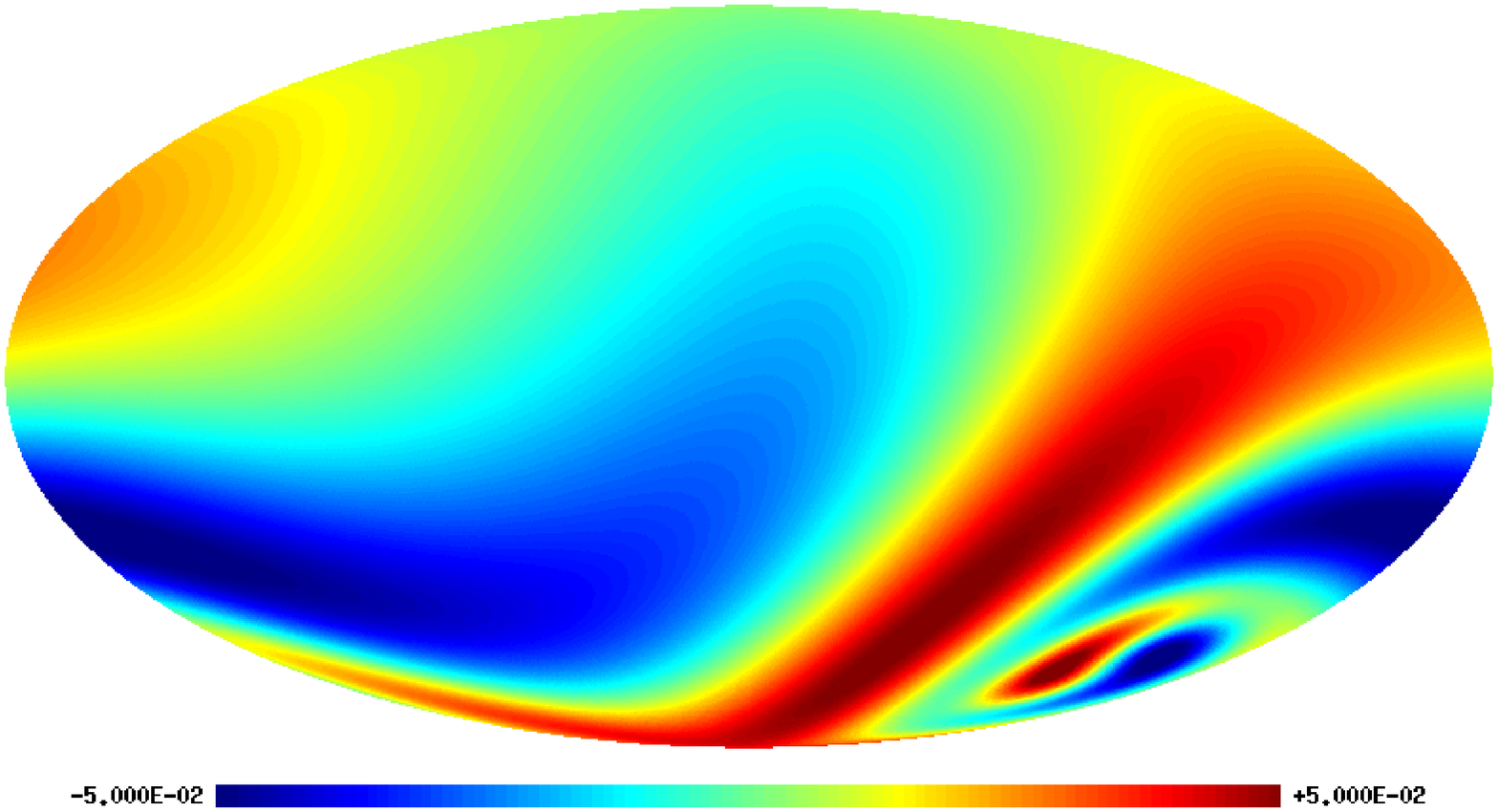}}
\caption{Full-sky CMB and correction maps displayed in the Mollweide projection.  The original WMAP ILC map is illustrated in panel (a), while the cold spot corrected map is shown in panel (b).  The cold spot template is displayed on a full-sky map in panel (c).  In panel (d) the best-fit \bianchiviih\ template determined by \citet{bridges:2006b} is shown.  Notice that the cold spot texture template aligns closely with the central cold swirl of the best-fit Bianchi template.  All maps are given in units of mK.}
\label{fig:maps}
\end{figure}

\begin{figure}
    	\subfigure{
          \includegraphics[width=.42\columnwidth]{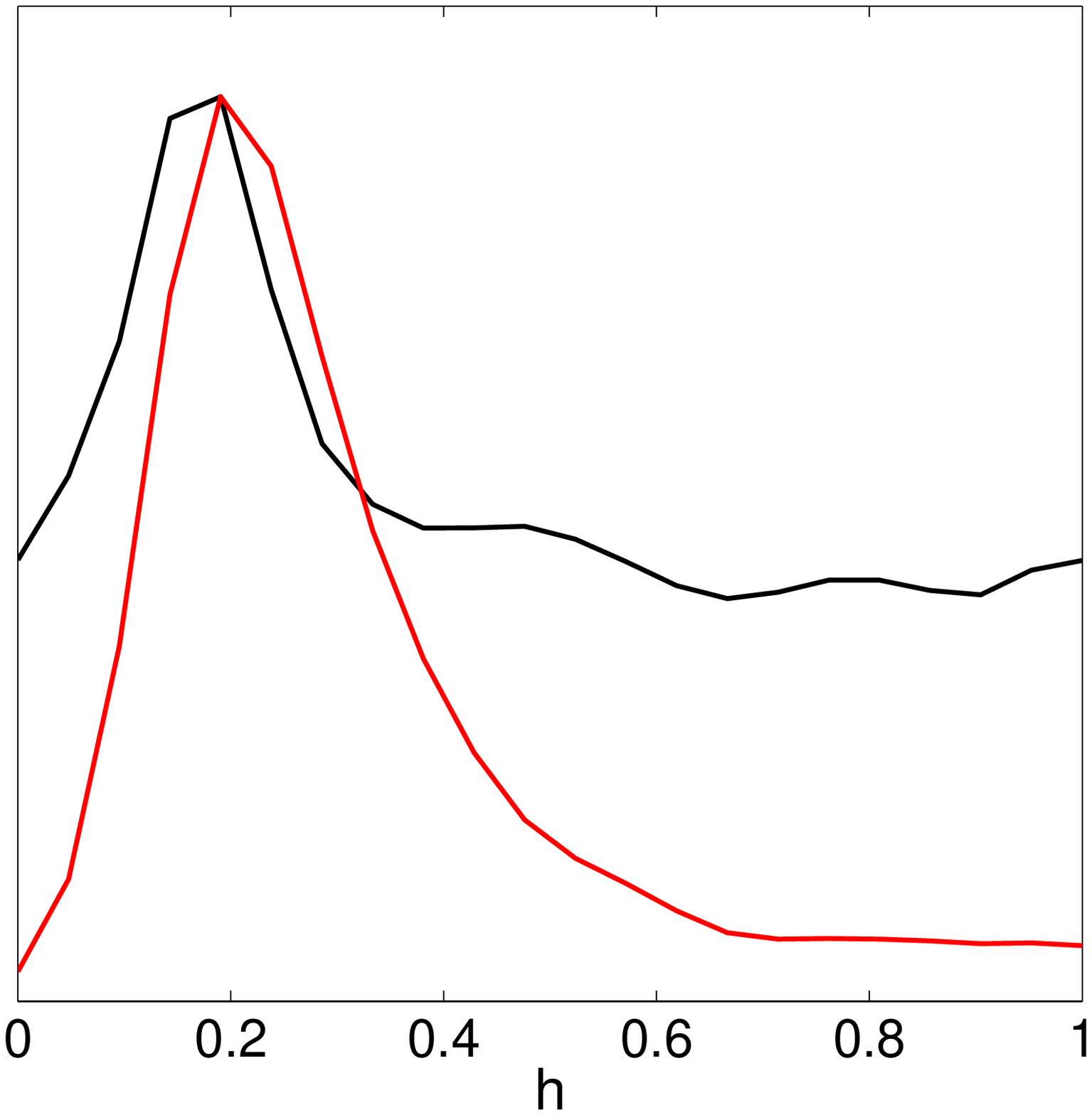}}
	\hspace{0.2cm}
	\subfigure{
          \includegraphics[width=.42\columnwidth]{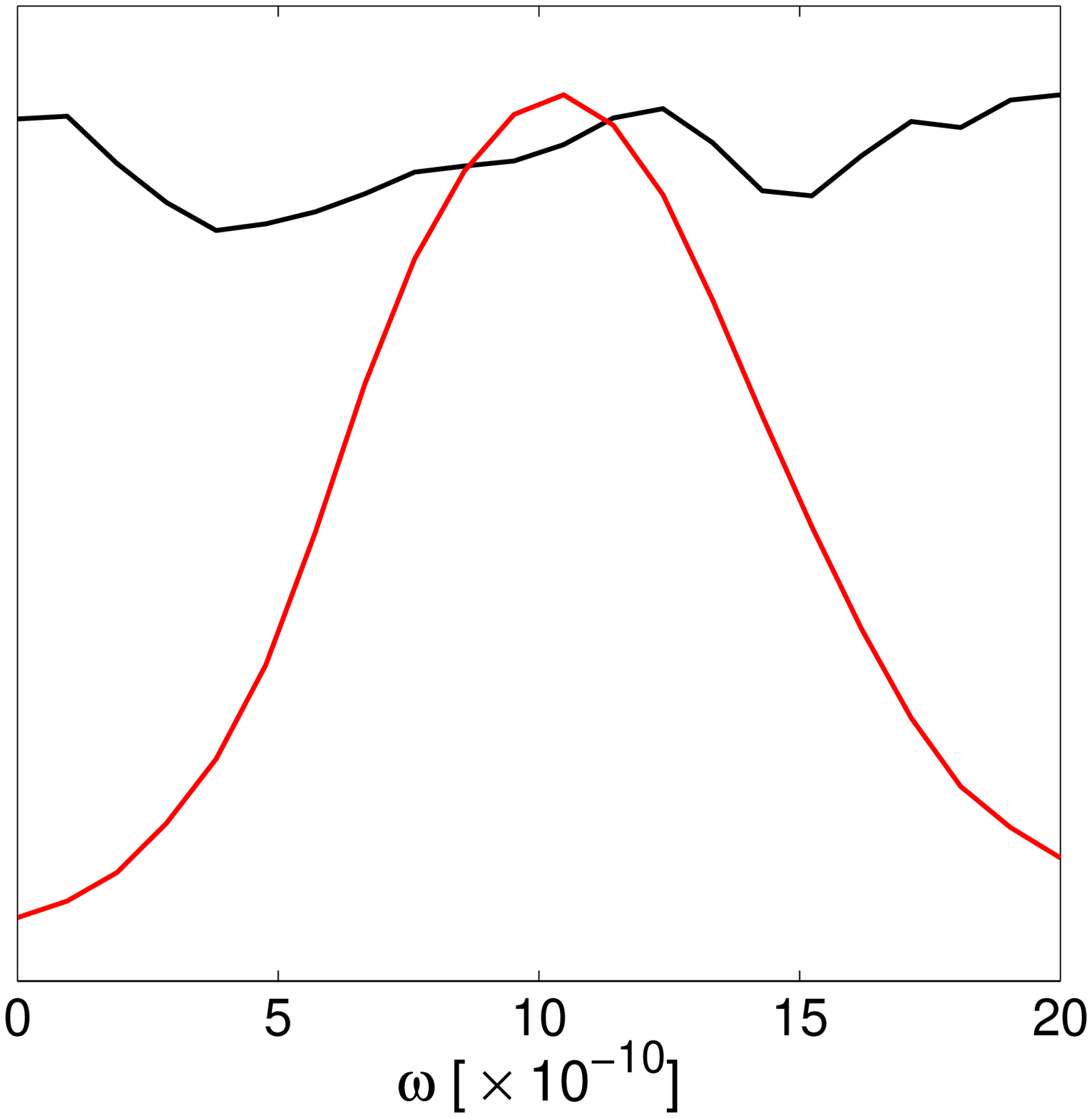}}\\
     	\subfigure{
           \includegraphics[width=.42\columnwidth]{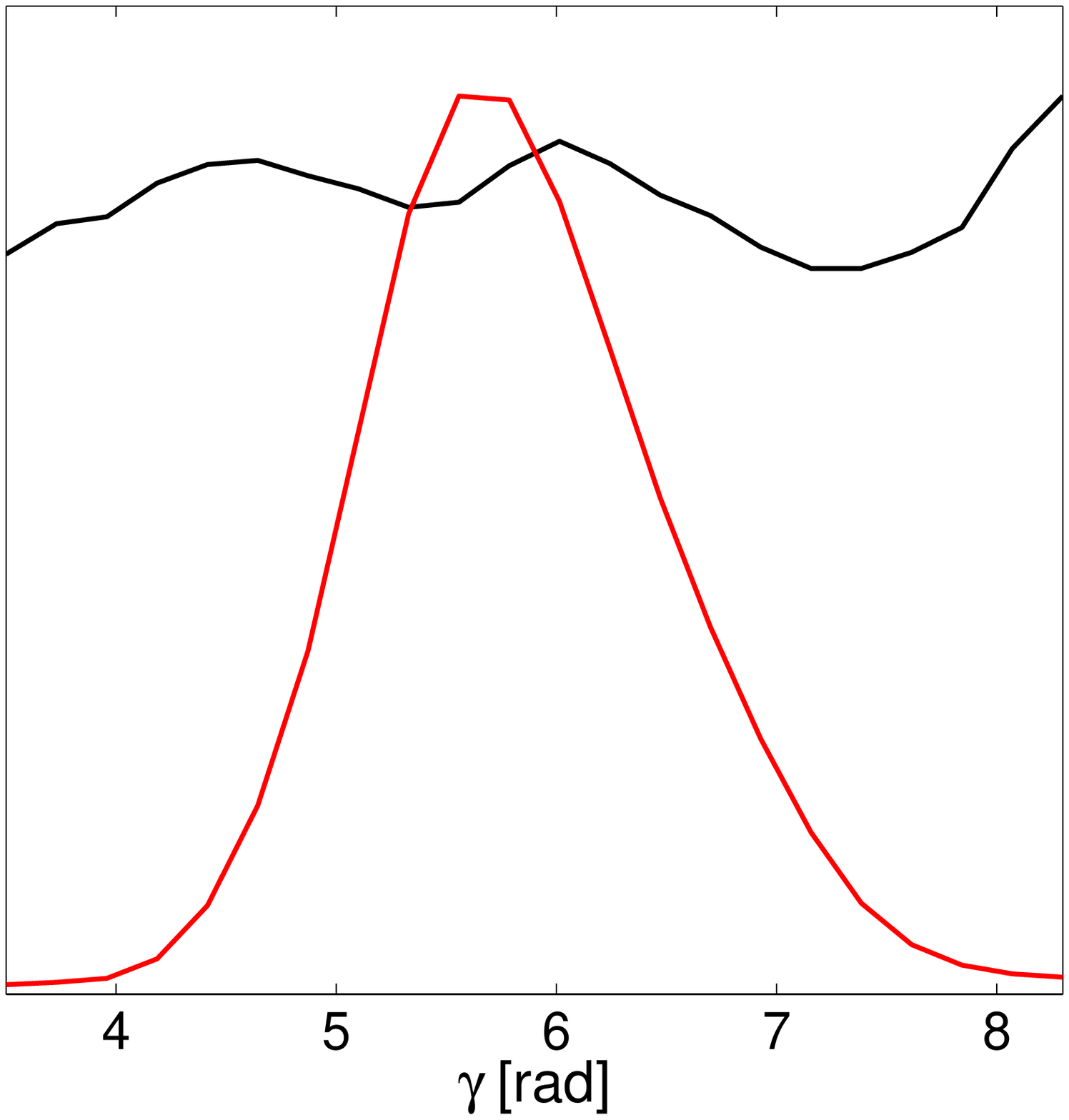}}
	\hspace{0.2cm}
	\subfigure{
	   \includegraphics[width=.42\columnwidth]{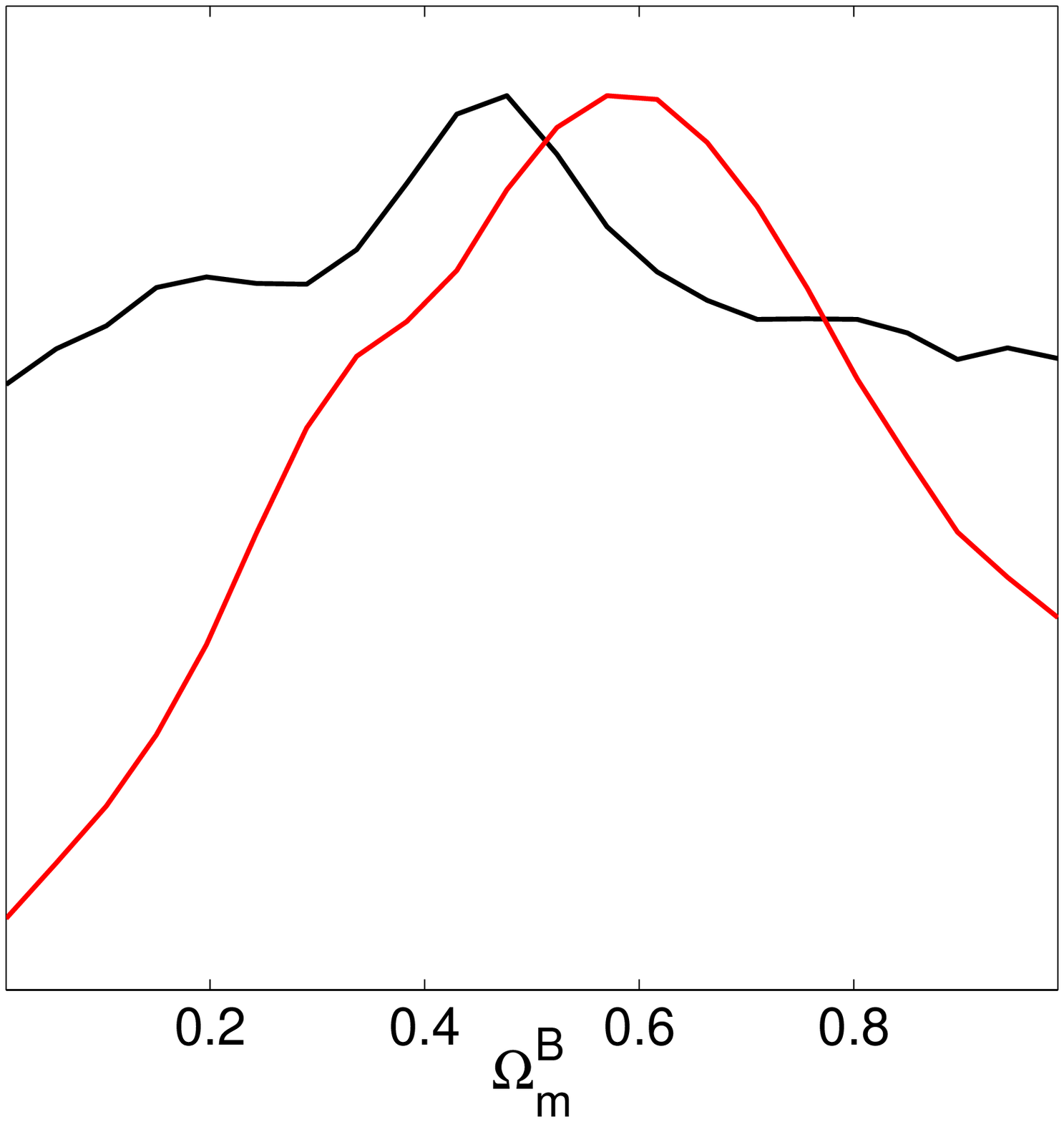}}\\
	\subfigure{
          \includegraphics[width=.42\columnwidth]{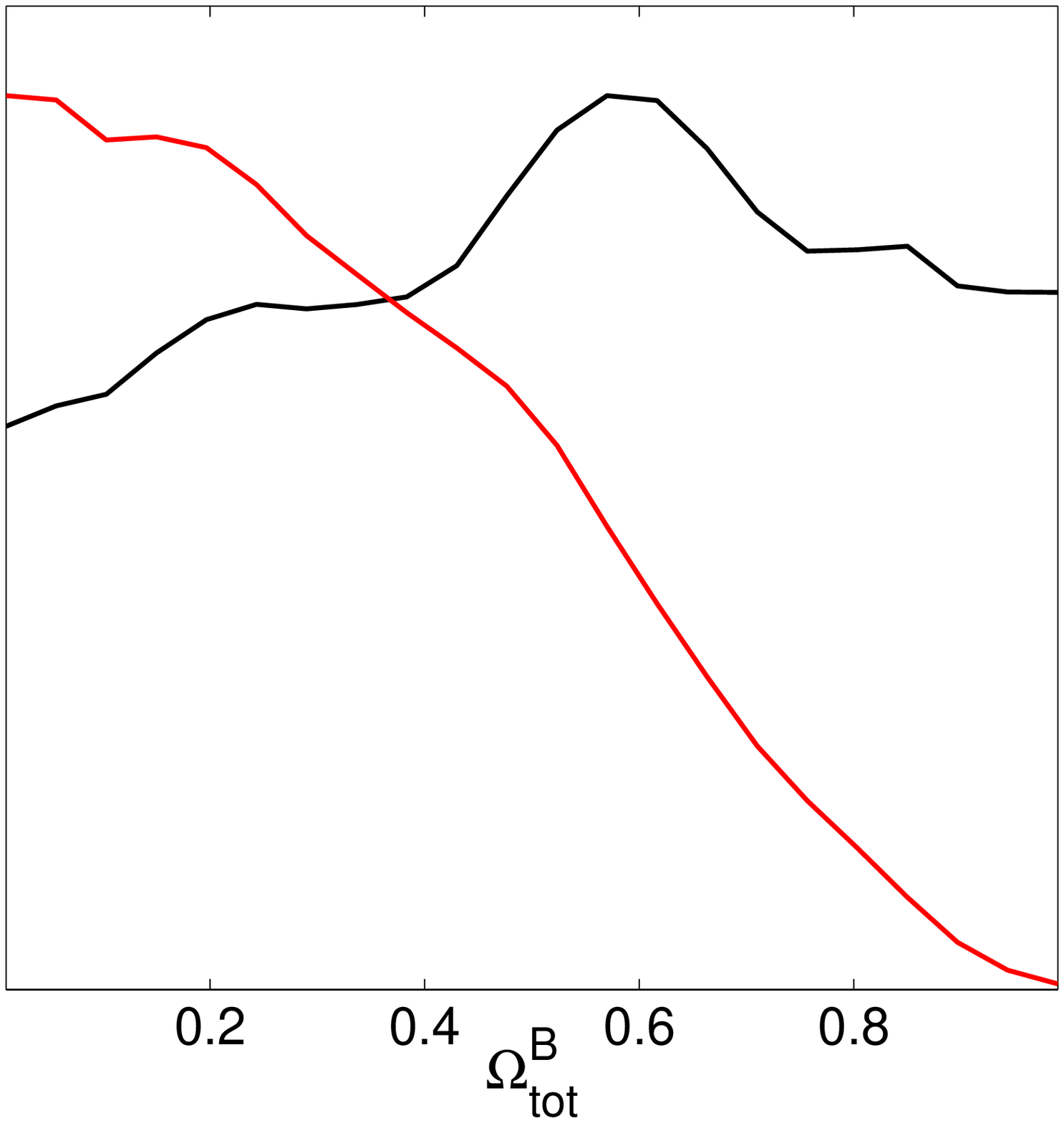}}   
\caption{Marginalised \bianchiviih parameters for left-handed model D using the original \wmap ILC map (red) and that corrected for the texture fit of
\citet{cruz:2007} (black).}
\label{fig:modelD}
\end{figure}

\begin{figure}
    	\subfigure{
          \includegraphics[width=.42\columnwidth]{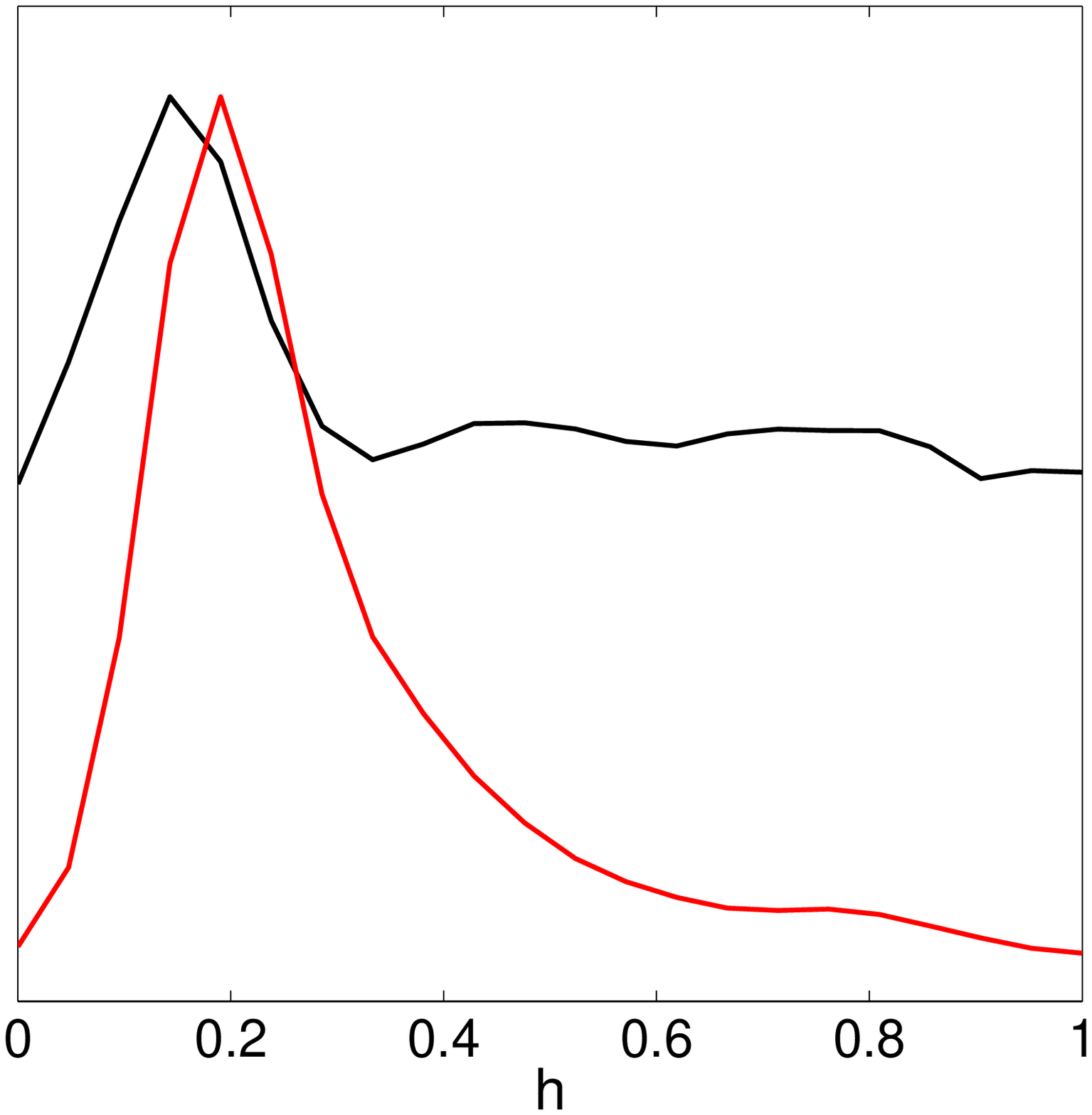}}
	\hspace{0.2cm}
	\subfigure{
          \includegraphics[width=.42\columnwidth]{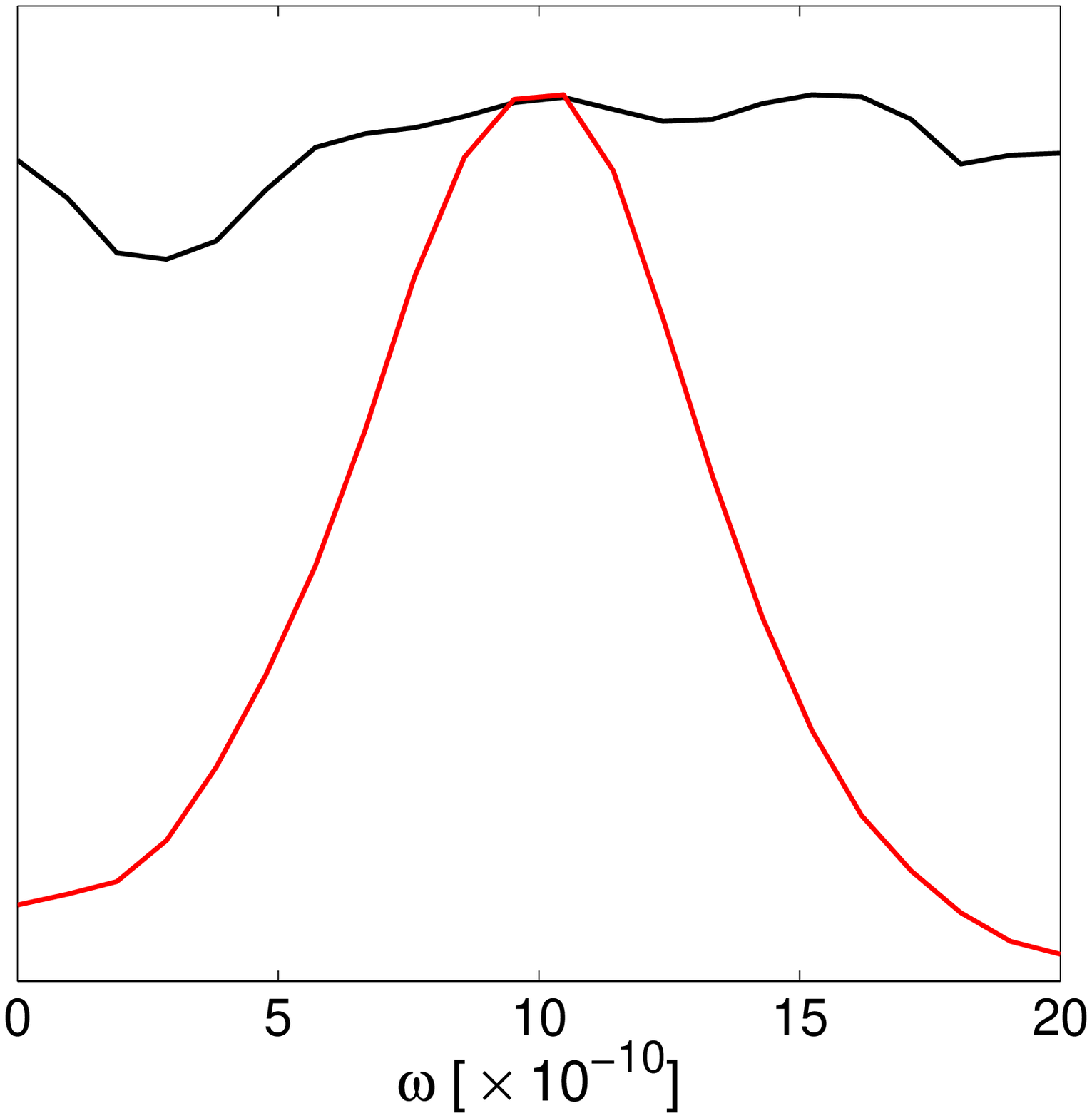}}\\
     	\subfigure{
           \includegraphics[width=.42\columnwidth]{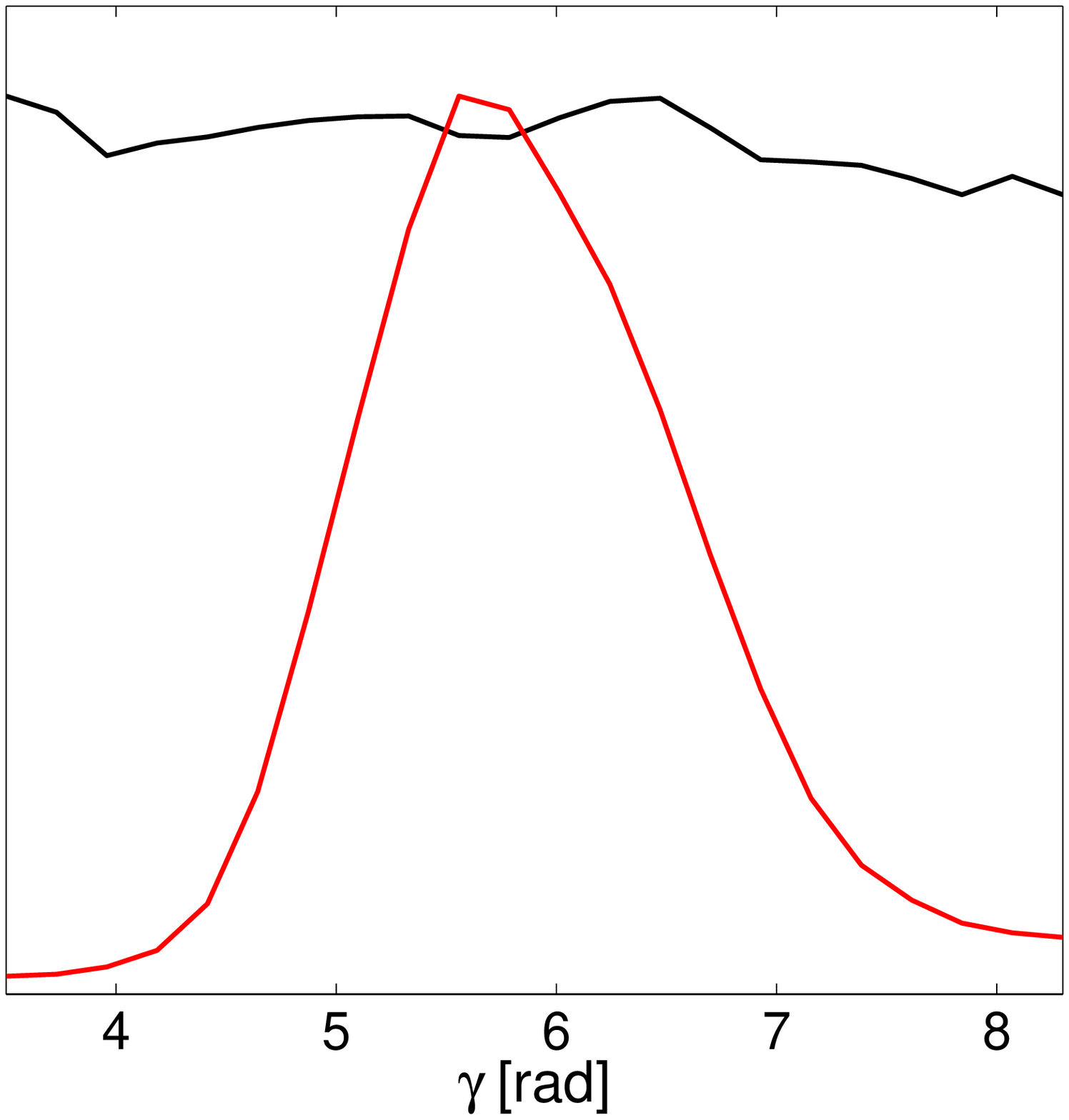}}
	\hspace{0.2cm}
	\subfigure{
	   \includegraphics[width=.42\columnwidth]{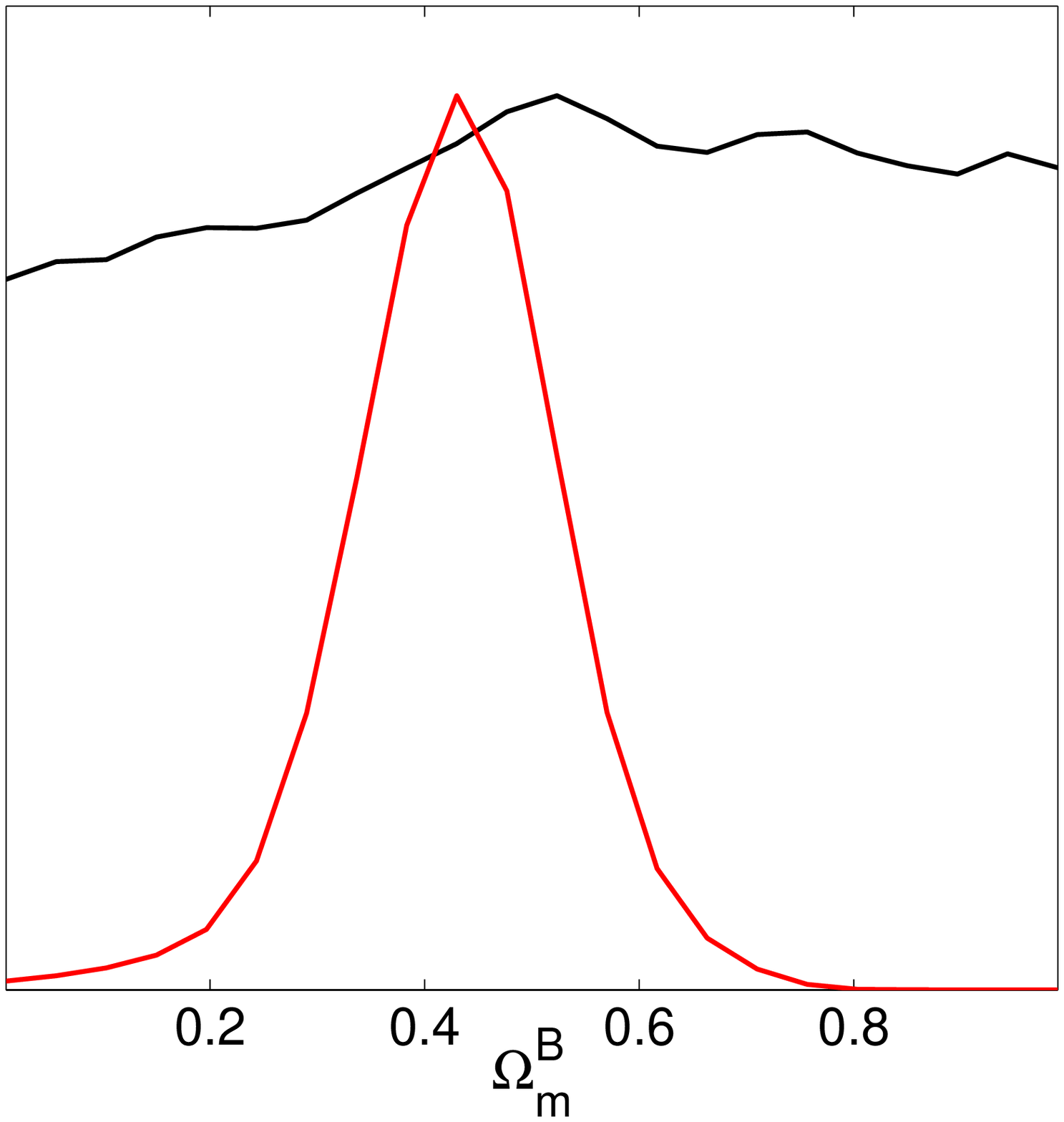}}
\caption{Marginalised \bianchiviih parameters for left-handed model G using the original \wmap ILC map (red) and that corrected for the texture fit of
\citet{cruz:2007} (black).}
\label{fig:modelG}
\end{figure}

In \citet{bridges:2006b} both models D and G resulted in highly constrained marginalised
posterior distributions (red lines of \fig{\ref{fig:modelD}} \& \ref{fig:modelG}) on each of
the \bianchiviih parameters. In this analysis  (black lines of \fig{\ref{fig:modelD} \&
\ref{fig:modelG}}) however  all parameters except perhaps a slight preference for $h \sim
0.2$, are entirely  unconstrained, and crucially, a non-zero likelihood is observed in the
vorticity at $\omega = 0$.  In the \bianchiviih formalism $\omega$ is highly correlated with
\bianchiviih signal amplitude so such a result illustrates that with the texture corrected
map the data no longer prefers any additional components of this kind over a standard
cosmology.  Although redundant, given the unconstrained posteriors, the same conclusions can
be confirmed via the Bayesian evidence estimates (see Table \ref{table:evidences}). Both
models D and G record significantly reduced evidences, model D is now about as favoured as a
$\Lambda$CDM cosmology alone (model A) while model G is now, almost significantly, \emph{disfavoured}
when compared with A. In \citet{bridges:2006b} all right-handed models were found to have
disfavouring evidences. It is worth noting that this result remains unchanged with the
texture corrected data. Both left and right handed models are now equally disfavoured.    

\begin{table}
\begin{center}
\hspace{5cm}
\caption{Bayesian evidence differences (logarithmic) between models A and left-handed D \& G for the original \wmap 3-year ILC map and that corrected
for the texture fit of \citet{cruz:2007}.}  
\begin{tabular}{|c||c|}
    \hline
 \textbf{Data} $\backslash$ \textbf{Model} 	& D (L)\\
    \hline
 3-year \wmap  					& +1.2 $\pm$ 0.2\\
 3-year \wmap (texture corrected) 		& -0.1 $\pm$ 0.2\\
     \hline
 \textbf{Data} $\backslash$ \textbf{Model} 	& G (L)	 	\\
 	\hline    
 3-year \wmap 					& +1.2 $\pm$ 0.2\\
 3-year \wmap (texture corrected)		& -0.8 $\pm$ 0.2\\
     \hline 
\end{tabular}
\label{table:evidences}
\end{center}
\end{table}

\section{Conclusions}
\label{sec:conclusions}

In \citet{bridges:2006b} we concluded that the preferred \bianchiviih template
was likely to be heavily influenced by the cold spot in the southern galactic
hemisphere. Accepting the recent texture fit  \citep{cruz:2007} essentially
removes this feature from the \wmap data. Although there were other
discernible features that appeared reduced following `correction' of the data
by the best-fit \bianchiviih template, these were mostly concentrated close to
the galactic plane where little confidence can be placed in the map quality.
The Bayesian evidence in favour of the inclusion of a \bianchiviih component,
was only marginally significant for just two of the models we had previously
considered. Using data `corrected' for the texture, both of these models are now left 
almost entirely unconstrained by the data and consequently register
disfavouring log evidence values. This result raises further doubts over the
relevance of a \bianchiviih explanation for any of the anomalous features seen
in the current \wmap data.   

\section*{Acknowledgements}


This work was carried out on the COSMOS UK National Cosmology Supercomputer at DAMTP, Cambridge and we would like to thank S.
V. Treviso for his assistance. JDM would like to acknowledge Clare College, Cambridge for a Research
Fellowship. MC thanks the Ministerio de Educaci\~Ron y Ciencia for a predoctoral FPU
fellowship. PV thanks an I3P contract from the CSIC. MC, PV and EMG thank the Ministerio de Educaci\~Ron y Ciencia project
AYA2007-68058-C03-02. Some of the results in this paper have been derived using the \healpix\ package \citep{gorski:2005}.  We
acknowledge the use of the \lambdaarchtext\ (\lambdaarch).  Support for \lambdaarch\ is provided by the NASA Office of Space Science.

\bibliographystyle{mymnras_eprint}
\bibliography{bib}

\label{lastpage}
\end{document}